%% file: main.tex
\PassOptionsToPackage{colorlinks=true,allcolors=blue,breaklinks=true}{hyperref}
\documentclass{iopjournal}

\usepackage{graphicx}
\usepackage{dcolumn}
\usepackage{bm}
\usepackage{amsmath, amssymb}
\usepackage{orcidlink}
\usepackage{siunitx}
\usepackage[ruled,vlined]{algorithm2e}
\usepackage{calc}

\usepackage{multirow}
\usepackage{array}
\usepackage{booktabs}
\usepackage{siunitx}
\usepackage{xspace}
\usepackage[breaklinks=true]{hyperref}
\usepackage[capitalize, noabbrev]{cleveref}
\crefname{algocf}{algorithm}{algorithms}
\Crefname{algocf}{Algorithm}{Algorithms}
\usepackage{orcidlink}
\usepackage{placeins}

\newcommand{\pt}{\ensuremath{p_\text{T}}}

\newcommand{\lslim}{L-GATr-slim\xspace}

\newcommand{\RR}{{\mathbb{R}}}

\mathchardef\mhyphen="2D
\newcommand{\II}{{\mathbb{I}}}

\newcommand{\lrp}[1]{{\left( #1 \right)}}
\newcommand{\lrb}[1]{{\left[ #1 \right]}}
\newcommand{\lrc}[1]{{\left\lbrace #1 \right\rbrace}}
\newcommand{\lra}[1]{{\left\langle #1 \right\rangle}}
\newcommand{\abs}[1]{{\left \vert #1 \right \vert}}
\newcommand{\norm}[1]{{\left\lVert #1\right\rVert}}

\newcommand{\EE}{\mathbb{E}}

\newcommand{\equald}{{\buildrel d \over =}}

\newcommand*\diff{{\mathop{}\!\mathrm{d}}}

\newcommand*\e{\mathop{}\!\mathrm{e}}
\newcommand*\imag{{\mathrm{i}}}

\DeclareSIUnit{\electronvolt}{e\kern-0.1em\relax{}V}
\DeclareSIUnit{\eV}{\electronvolt}

\renewcommand{\articletype}[1]{%
  \vspace*{4mm}\noindent{\scriptsize\sf{\bfseries \MakeUppercase{#1}}}\par\vspace{2mm}%
}
\renewcommand{\headrulewidth}{0pt}

\begin{document}

\articletype{Paper}

\title{Lorentz symmetry in self-supervised representation learning}

\author{Andreas Hermansen\(^{1,*}\)\orcidlink{0009-0006-1162-9770}, Tobias Golling\(^1\)\orcidlink{0000-0001-8535-6687} and Slava Voloshynovskiy\(^{2}\)\orcidlink{0000-0003-0416-9674}}

\affil{$^1$Département de physique nucléaire et corpusculaire, University of Geneva, Geneva, Switzerland}

\affil{$^2$Department of Computer Science, University of Geneva, Geneva, Switzerland}

\affil{$^*$Author to whom any correspondence should be addressed.}

\email{andreas.hermansen@unige.ch}

\input{body}
\end{document}

%% file: body.tex
\begin{abstract}
Self-supervised learning (SSL) and explicit physical inductive biases have driven major advances in jet tagging, yet they are typically pursued along separate tracks. In this work, we unify both paradigms by integrating the Lorentz-equivariant Geometric Algebra Transformer slim (\lslim) architecture with the LeJEPA framework for representation learning. Pre-training on 100 million jets from the JetClass dataset, we optimize towards a latent space, which is invariant across physics-motivated jet augmentations. We demonstrate that enforcing Lorentz symmetry in the latent space is critical for representation quality and that this can be achieved either architecturally or learned via kinematic augmentations. With a standard non-invariant Transformer backbone, LeJEPA outperforms established autoregressive, masked particle modelling, and self-distillation baselines in linear- and MLP-probe accuracy, with \lslim providing further complementary gains. Across downstream benchmarks, our pre-trained models consistently outperform baselines trained from scratch or through different pre-training objectives.
\end{abstract}

\section{Introduction}
\label{sec:intro}
In high energy physics (HEP), increasingly sophisticated machine learning algorithms are continually being developed. In jet tagging, this progress has followed two largely disjoint tracks: self-supervised learning (SSL) and the incorporation of physical inductive bias, in particular through Lorentz-invariant architectures.

The development of SSL in HEP has been strongly influenced by its success in broader domains such as natural language processing (NLP) and computer vision (CV). 

In NLP, reconstruction-based approaches, specifically token prediction in the discrete input space, remain the dominant paradigm. This methodology has evolved along two trajectories: autoregressive next-token prediction~\cite{gpt, gpt3, deepseek4} and non-causal masked-token prediction~\cite{bert, bart, llada21}. Although these approaches appear distinct, they are closely related mathematically~\cite{radd, anyorder}.

In CV, reconstruction-based approaches have also been explored~\cite{mae, DAE}, but the field has shifted towards representation-based approaches, which aim to directly construct a semantically meaningful latent representation. Early work in this direction relied on contrastive learning, in which similar samples are pulled together in latent space and dissimilar samples are pushed apart~\cite{simclr}. More recent methods have moved away from contrastive learning in favour of approaches that operate purely on similar samples, such as self-distillation~\cite{dino, dinov3} and joint-embedding predictive architectures (JEPAs)~\cite{ijepa, lejepa}, both of which rely on various regularisation schemes to prevent representation collapse.

Motivated by these developments, numerous adaptations of SSL to HEP have been explored. These span the full methodological spectrum: from reconstruction-based approaches, such as masked particle modelling~\cite{mpm, mpm2}, autoregressive next-token prediction~\cite{birk2024omnijetalpha}, and mixtures thereof~\cite{birk2026omnijetalpha2}, to representation-based approaches like contrastive learning~\cite{JetCLR, harris2024rs3l}, self-distillation~\cite{rino}, and JEPAs~\cite{jjepa, hepjepa, jpjepa}.

In parallel, the incorporation of physical inductive bias into neural networks has emerged as a highly successful direction in various HEP tasks such as generative modelling, QFT amplitude surrogates, and jet tagging~\cite{lgatr, lloca}. These approaches fall into two broad categories. The first includes physically motivated invariants, such as the pairwise invariant mass, directly into the model inputs~\cite{parT, mipart}. The second achieves exact Lorentz invariance, either through the network architecture~\cite{lgatr, lloca, lorentznet, lgatr2,  lloca2, lgatrslim} or by operating exclusively on invariant features~\cite{pelican}. Both of these latter approaches break the full symmetry to better match the detector environment. Such architectures impose a strong constraint on the admissible function class of the neural network, thus preventing it from becoming non-physical. 

Recent work has compared encoding symmetries in the architecture with learning them through SSL, treating the two as alternative strategies~\cite{explicitandimplicit}. In this work, we instead combine the two approaches by integrating Lorentz-invariant architectures with the representation-based SSL method LeJEPA~\cite{lejepa}.

Our code is available at \texttt{\href{https://github.com/AMHermansen/lijepa}{https://github.com/AMHermansen/lijepa}}

\section{Related work}
\label{sec:related}
While JEPAs have been explored in the context of HEP in the form of J-JEPA~\cite{jjepa}, HEP-JEPA~\cite{hepjepa}, and JetParticle-JEPA~\cite{jpjepa}. However, these applications have exclusively centred around the I-JEPA framework~\cite{ijepa}, which relies on predicting the representations of masked-out portions of a jet from an unmasked context. To avoid representation collapse, these methods use a student-teacher setup equipped with exponential moving average (EMA) synchronisation of the model parameters and stop-gradient operators. 

In this work, we instead opt for the LeJEPA formulation~\cite{lejepa}. Rather than focussing on masked patch prediction, this approach aims to produce a latent representation invariant under different views, which are obtained by applying various augmentations to the data. Consequently, this methodology is much closer in spirit to contrastive learning and self-distillation approaches previously established in HEP. For instance, JetCLR~\cite{JetCLR} aligns augmented views of the same jet using a contrastive objective, while RINO~\cite{rino} uses a self-distillation approach. Re-simulation-based self-supervised representation learning (RS3L)~\cite{harris2024rs3l} also enforces representation invariance through a contrastive training objective. However, it relies on paired re-simulations of the same physical event rather than augmentations, making the method less suited for purely data-driven SSL on real detector data.

\section{Data}
\label{sec:data}
A key motivation for SSL in HEP is the prospect of eventually training neural networks directly on unlabelled detector data. While this has been explored in Reference~\cite{aspen}, performance evaluation still relies on simulated data. As such, we utilise the JetClass dataset~\cite{jetclass} published in conjunction with Reference~\cite{parT} for pre-training. Although this dataset contains classification labels, our pre-training objective ignores them. Thus, our method is directly transferable to real detector data.

\subsection{Datasets}
\paragraph{JetClass}
The JetClass dataset~\cite{jetclass} originally released with Reference~\cite{parT} serves as our primary pre-training and in-domain dataset. It consists of 125 million simulated large-R jets distributed evenly across 10 distinct classes. The physical processes are generated using \textsc{MadGraph5\_aMC@NLO}~\cite{madgraph} with parton showering and hadronisation performed by \textsc{Pythia8}~\cite{pythia}. The detector response is simulated in \textsc{Delphes}~\cite{deFavereau:2013fsa} using the default CMS detector card. Constituents are clustered into jets using the anti-\(k_t\) algorithm~\cite{antikt} with a distance parameter of \(R = 0.8\). 

In a basic kinematic selection reconstructed jets are required to have transverse momenta of \(\qty{500}{\giga\electronvolt} < p_\text{T} < \qty{1000}{\giga\electronvolt}\) and pseudorapidities of \(|\eta| < 2.0\). We use the provided dataset partition into 100 million jets for training, 5 million for validation, and 20 million for testing. Additionally, we follow standard practice on JetClass and restrict all jets to 128 constituents. In addition to quantities derived from four-momenta, the dataset also includes longitudinal impact parameter and a corresponding uncertainty \((z_0, \sigma(z_0))\) and likewise for the transverse impact parameter \((d_0, \sigma(d_0))\) for all charged particles. The impact-parameter related parameters are zero-padded for neutral particles. In addition to these parameters, 8-dimensional one-hot encodings of a combination of charge / photon / hadron information is also available. A concrete description of this encoding is detailed in \cref{tab:id_variables}.

\begin{table}[t]
    \centering
    \caption{Summary of identification variables in JetClass.}
    \label{tab:id_variables}
    \begin{tabular}{l r}
        \toprule
        Particle description & Identification \\
        \midrule
        Photon & 1 \\
        Positive hadron & 2 \\
        Negative hadron & 3\\
        Neutral hadron & 4\\
        Electron & 5\\
        Positron & 6\\
        Muon & 7\\
        Anti-muon & 8\\
        \bottomrule
    \end{tabular}
\end{table}

\paragraph{Top tagging reference dataset}
To benchmark the transferability of the learned representations, we evaluate downstream performance on the top tagging dataset~\cite{toptag_zenodo} published together with Reference~\cite{toptag_paper}. This dataset provides a standardised binary classification task to discriminate highly boosted top quarks from light quark and gluon jets. It contains 2 million jets generated at a centre-of-mass energy of \(\sqrt{s} = \qty{14}{\tera\electronvolt}\) using \textsc{Pythia8}~\cite{pythia}, with detector effects similarly modelled by the \textsc{Delphes}~\cite{deFavereau:2013fsa} fast simulation using the default ATLAS detector card. The jets are clustered using the anti-\(k_t\) algorithm~\cite{antikt} (\(R=0.8\)) and are subjected to a kinematic selection of \(\qty{550}{\giga\electronvolt} < p_\text{T} < \qty{650}{\giga\electronvolt}\) and \(|\eta| < 2.0\). We use the provided dataset split into \num{1200000} training samples, \num{400000} validation samples, and \num{400000} testing samples. In contrast to the JetClass dataset, this dataset purely contains the four-momenta.

In summary, across both datasets each jet is represented as an unordered set of \(N\) constituent particles, \(X = \lrc{x_i}_{i=1}^N = \lrc{\lrp{u_i, s_i}}_{i=1}^N\). Each constituent \(x_i\) comprises a four-momentum \(u_i \in \RR^{1,3}\) that transforms equivariantly under the Lorentz group together with optional scalar features \(s_i \in \RR^{d_s}\) (with \(d_s = 0\) for the top tagging dataset) which are treated as Lorentz-invariant.

\section{\label{sec:method} Method}

\subsection{\label{sec:architecture} Architecture}
To isolate the impact of Lorentz symmetry during self-supervised pre-training, we use a standard Transformer as our non-invariant baseline backbone and compare it against the Lorentz-equivariant Geometric Algebra Transformer slim (\lslim) architecture~\cite{lgatrslim}. Because \lslim closely mirrors a standard Transformer, performance differences are more directly attributable to Lorentz-invariance.

\lslim operates natively on four-momenta and Lorentz scalars through separate vector and scalar channels, allowing us to directly feed in the lab-frame constituent four-momenta \(u_i\) and scalar features \(s_i\). For the standard Transformer, we instead follow the kinematic input pre-processing of the Particle Transformer (ParT)~\cite{parT}. To account for the symmetry breaking of collider events relative to the beam axis, we append constant reference beam-axis and time spurion tokens to the \lslim input. To ensure the non-invariant baseline has identical sequence capacity, we append a matching number of learnable register tokens to the standard Transformer input, which have independently been shown to improve attention representations in NLP~\cite{attentionsink} and CV~\cite{registers} models.

After passing the constituent sequence through the primary backbone, we map the per-constituent outputs into a \(d\)--dimensional feature space prior to sequence pooling. For the standard Transformer, we retain all latent constituent features directly. For \lslim, we construct purely Lorentz-invariant constituent features by concatenating the latent scalars, which are Lorentz-invariant, with the Minkowski inner products of the latent vectors. Due to our choice of hyperparameters, both approaches yield the exact same latent dimensionality \(d\). Because the resulting latent representations are Lorentz-invariant in \lslim, we aggregate the sequence in both architectures by prepending a learnable \textsc{cls} token, passing the set through two standard Transformer layers, and extracting the final \textsc{cls} output as the jet representation \(z \in \RR^d\). For \lslim this results in a single Lorentz-invariant embedding vector, whereas the Transformer produces a general representation without any symmetry guarantees.

\subsection{\label{sec:lejepa} LeJEPA}
\begin{figure}[t]
    \centering
    \includegraphics[width=0.3\linewidth]{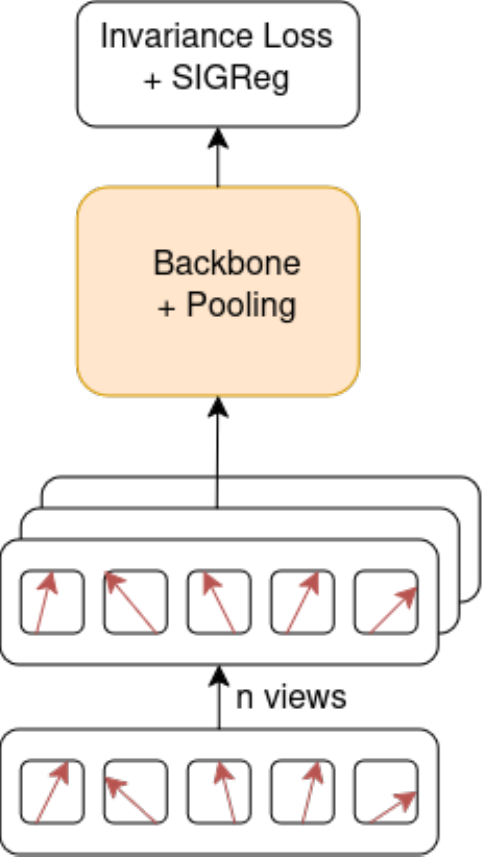}
    \caption{Overview of the LeJEPA pre-training pipeline for jet representation learning. For each jet, \(n\) stochastic views are generated via data augmentations. Each view is processed by the backbone encoder and sequence-pooling stage to produce a latent representation \(z_{b,v} \in \RR^d\), which is optimised via a combination of an invariance loss described in \cref{eq:invariance} and the SIGReg distributional regulariser described in \cref{eq:sigreg_expectation}.}
    \label{fig:lejepa_framework}
\end{figure}

In \cref{fig:lejepa_framework} we schematically illustrate the LeJEPA pre-training framework, in which the objective~\cite{lejepa} is defined as a weighted combination of a view-invariance loss and a distributional regularizer:
\begin{equation}
    \mathcal{L}_\text{LeJEPA} = \lrp{1 - \lambda} \mathcal{L}_\text{invariance} + \lambda \mathcal{L}_\text{SIGReg},
\end{equation}
where \(\lambda \in (0, 1)\) balances the two terms. Given a mini-batch of jets, \(B\), we generate a set of stochastic views \(V_b\) for each jet \(b \in B\) (see \cref{sec:augmentations} for the exact procedure). We then compute their projected embeddings \(z_{b,v} = f_\theta(X_{b,v}) \in \RR^d\), where \(b \in B\) and \(v \in V_b\). The invariance term is responsible for pulling together similar events in latent space. This is achieved by pulling together different views of the same event through
\begin{equation}\label{eq:invariance}
    \mathcal{L}_\text{invariance} = \frac{1}{\abs{B}} \sum_{b \in B} \frac{1}{\abs{V_b}}\sum_{v \in V_b} \norm{z_{b,v} - \bar{z}_b}_2^2,
\end{equation}
where \(z_{b, v}\) denotes the latent representation of jet \(b\) with view \(v\) and \(\bar{z}_b = \frac{1}{\abs{V_b}} \sum_{v \in V_b} z_{b, v}\) denotes the empirical mean computed over different views. We note that training a model to be invariant under Lorentz-transformations can be viewed as a Symmetry EncourAging Loss (SEAL)~\cite{seal}, but where SEAL includes this as an additional auxiliary term, it is the primary learning objective of LeJEPA.

To prevent representation collapse, LeJEPA~\cite{lejepa} introduced the Sketched Isotropic Gaussian Regularization (SIGReg) term, which shapes the empirical distribution of latent embeddings \(Z = \lrc{z_b}_{b \in B}\) to match a pre-specified target distribution. While LeJEPA uses a standard multivariate normal distribution as their target, subsequent works have explored uniform hyperspherical distributions~\cite{spherejepa, spherejepa2} and multivariate Laplace distributions~\cite{kerjepa}. In our setting, regularising toward a multivariate Laplace distribution yields the best probe performance (see \cref{app:ablation:sigreg} where we include results for the normal and hyper-spherical regularisation target).

By the Cram\'{e}r-Wold theorem, two random variables \(Y, Y' \in \RR^d\) have identical distribution \(Y \equald Y' \) if and only if all of their one-dimensional marginal projections have identical distributions \(\lra{Y, a} \equald \lra{Y', a}\) along every \(d\)--dimensional unit direction \(a \in S^{d-1}\)~\cite{lejepa, cramerwold1936}. SIGReg relies on this theorem to measure the discrepancy between a mini-batch of embeddings \(Z\) and the target distribution via a one-dimensional statistical test over multiple random projections instead of directly measuring the discrepancy in the high-dimensional embedding space. Following References~\cite{lejepa, spherejepa}, we use the Epps-Pulley test~\cite{eppspulley1983} as our one-dimensional statistical test. As a result our SIGReg loss becomes
\begin{equation}\label{eq:sigreg_expectation}
    \mathcal{L}_\text{SIGReg} = \EE_{a} \lrb{\int_{-\infty}^\infty \abs{\hat{\phi}_{\lra{Z, a}}(s) - \phi(s)}^2 w(s) \, \diff s},
\end{equation}
where the expectation is taken over the normalised uniform measure on \(S^{d-1}\), \(w(s) = \exp(-s^2/2)\) is a Gaussian spectral weighting function, and \(\phi(s)\) is the characteristic function of the one-dimensional marginal distribution of the target. Denoting the projected mini-batch along \(a\) by \(\lra{Z, a} = \lrc{\lra{z_b, a}}_{b \in B}\), the corresponding empirical characteristic function \(\hat{\phi}_{\lra{Z, a}}(s)\) is given by
\begin{equation}
    \hat{\phi}_{\lra{Z, a}}(s) = \frac{1}{\abs{B}} \sum_{b \in B} \e^{\imag s \lra{z_b, a}}.
\end{equation}
For each mini-batch, \cref{eq:sigreg_expectation} is estimated by random sampling of multiple \(a\), and the integral is evaluated numerically using the trapezoid rule and 17 knots over the interval \(\lrb{-3, 3}\). For a more complete description of the LeJEPA method, we refer to Reference~\cite{lejepa}.

\subsection{\label{sec:augmentations} Generating views}
In order to generate different views of a jet, we use augmentations based on the prior work of JetCLR~\cite{JetCLR} and RINO~\cite{rino}. Our collection of views always include the original unmodified jet as one of the views.

\paragraph{JetCLR}
For the JetCLR based augmentations we first apply collinear splitting. This is done by randomly selecting \(\qtyrange[range-phrase=\textendash]{10}{50}{\percent}\) of the particles in a jet and then splitting each into two new particles carrying \((1-f)\) and \(f\) of the 4-momentum of the initial particle, with \(f\) sampled uniformly on the unit interval. The impact parameters and the identification information are directly copied from each split constituent to its daughters.
Afterwards, we smear the relative pseudo-rapidity \(\Delta \eta\) and azimuthal angle \(\Delta \phi\) compared to the jet axis. That is we sample 
\begin{equation}
    \lrp{\Delta \eta', \Delta \phi'} \sim \mathcal{N}\lrp{\lrp{\Delta \eta, \Delta \phi}, \frac{\II}{\max\lrp{\pt / \unit{\giga\electronvolt}, 0.1}}}.
\end{equation}
Beyond this we also randomly sub-sample each jet by randomly discarding \(30\unit{\percent}\) of the particles.

\paragraph{RINO}
For the RINO-based augmentations we use the \(k_\text{T}\) clustering implemented through \textsc{FastJet}~\cite{fastjet}. While Reference~\cite{rino} creates both global and local views using this algorithm, we instead treat all views as global. This, makes the two augmentation approaches methodologically more similar since there are no confounding differences by having local views in addition to global views. In total we sample eight views containing \(\lrc{1, 2, 3, 4, 8, 16, 32, 64}\) sub-jets.

Beyond these base augmentations, we also experiment with applying global Lorentz boosts and spatial rotations across all generated views. In both architectures, this encourages the model to learn representations that are invariant under these kinematic transformations. We note that \lslim the model is not fully invariant under full Lorentz-transformations due to the incorporation of spurion tokens. Therefore, Lorentz transformations can contribute to the invariance loss. 

\section{Results}
\label{sec:results}
For evaluation of different pre-training methods, we report the linear and MLP-probe accuracy on JetClass. These probes measure the accuracy of a linear- or MLP-classifier trained on the frozen latent representation. We obtain these metrics by jointly training the probe heads during pre-training on the frozen embeddings. After pre-training we fine-tune the probe heads on \num{1e6} jets using full-batch L-BFGS~\cite{nocedal1980lbfgs}. These two metrics thus capture how feature rich the produced latent space is. Freezing a foundation model such that point cloud inputs can be mapped to \(\RR^d\) would enable these high-dimensional latent representations to be used directly in further analyses by calibrating the latent space~\cite{calibration}. Beyond these probes, we evaluate the performance of full fine-tuning of the pre-trained model on limited training samples both in domain on JetClass and out of domain on the top tagging dataset.

\subsection{\label{sec:res:views} Dependence on views and Lorentz invariance}
To disentangle the impact of explicit architectural symmetry from that of data-driven invariance, we evaluate both the standard Transformer and \lslim across the JetCLR and RINO view-generation strategies outlined in \cref{sec:augmentations}. Additionally we evaluate augmentations both containing and excluding global \(\text{SO}^+(1,3)\) Lorentz transformations i.e. boosts and rotations. Because the reference spurions in \lslim explicitly break full \(\text{SO}^+(1,3)\) symmetry down to the circular subgroup consisting of rotations azimuthal around the beam axis, we also evaluate a strictly \(\text{SO}^+(1,3)\)-invariant \lslim model by removing the beam and time spurions. The resulting linear- and MLP-probe accuracies evaluated on JetClass are presented in \cref{tab:result_augmentation_probes}.

\begin{table}[t]
    \centering
    \caption{Comparison of linear- and MLP-probe accuracies evaluated on JetClass across view-generation strategies and backbone architectures. Here, \(+\,\text{SO}^+(1,3)\) denotes the addition of random global Lorentz boosts and spatial rotations to the base augmentations, as described in \cref{sec:augmentations}. The Transformer model incorporates no physical inductive bias, while \lslim is invariant to azimuthal rotations around the beam axis and \lslim (no spurions) is fully invariant under \(\text{SO}^+(1,3)\) (up to machine precision).}
    \label{tab:result_augmentation_probes}
    \begin{tabular}{llcccc}
    \toprule
    & & \textit{RINO} & \textit{RINO} & \textit{JetCLR} & \textit{JetCLR} \\
    & &               & \(+\,\text{SO}^+(1,3)\) &       & \(+\,\text{SO}^+(1,3)\) \\
    \midrule
    \multirow{3}{*}{Linear}
        & Transformer              & 0.414 & 0.732 & 0.331 & 0.762 \\
        & \lslim                   & 0.457 & 0.741 & 0.480 & 0.774 \\
        & \lslim (no spurions)     & 0.763 & 0.763 & 0.763 & 0.772 \\
    \midrule
    \multirow{3}{*}{MLP}
        & Transformer              & 0.516 & 0.771 & 0.457 & 0.792 \\
        & \lslim                   & 0.525 & 0.774 & 0.586 & 0.800 \\
        & \lslim (no spurions)     & 0.794 & 0.793 & 0.803 & 0.806 \\
    \bottomrule
\end{tabular}
\end{table}

The results in \cref{tab:result_augmentation_probes} demonstrate that obtaining Lorentz invariance in the latent representation is an important feature of representation quality in LeJEPA. When trained exclusively with base augmentations without \(\text{SO}^+(1,3)\) transformations, both the standard Transformer and the spurion-equipped \lslim suffer severe performance degradation. To further illustrate the role of Lorentz invariance, we observe that \lslim without spurions, does not degrade in performance when the Lorentz transformations are excluded from the augmentations. We note that while mathematically \lslim without spurions should reach the exact same metrics for the augmentation pairs differing purely by inclusion of Lorentz transformations, this is not the case  due to finite precision of floating point numbers, in which different rounding errors can occur for different augmentation sets. This ultimately yield slightly different optimisation trajectories. We observe that the fully Lorentz-invariant architecture achieves the best performance when excluding Lorentz-transformations from the augmentations. When the non-fully invariant architectures are trained towards a fully invariant latent space, by the inclusion of Lorentz-transformations in the augmentations, they manage to recover most of the performance.

\subsection{\label{sec:res:ssl} Comparison to other SSL methods}
We compare the LeJEPA pre-training formulation against an autoregressive pre-training approach, a masked particle modelling approach, and a self-distillation approach.

\paragraph{Autoregressive approach}
For the autoregressive baseline, we follow a setup similar to Reference~\cite{birk2026omnijetalpha2}, embedding all input features continuously, applying Rotary Position Encoding (RoPE)~\cite{rope}, using causal attention, and pre-trained with a next-token-prediction (NTP) objective. We evaluate pre-training objectives over different subsets of constituent features: VQ-VAE~\cite{vqvae} discretised kinematics trained via cross-entropy, identification (ID) one-hot encodings trained via cross-entropy, and continuous impact parameters trained via \(L_2\) regression. Because the model is causal, we extract the final constituent token embedding as the jet representation for probing. This is the only constituent token with full event attention context enabling it to be used as a global representation. We note that regressing continuous impact parameters via an \(L_2\) objective invalidates direct autoregressive sampling. Additionally jointly predicting different discrete features with cross-entropy implicitly assumes that these features are conditionally independent, which further precludes the model from being used generatively. Despite these architectural choices, the model remains valid for representation learning. Across our ablations (\cref{app:ablation:ar}), we found that predicting purely the VQ-VAE tokens yielded the best probe performance. We note, however, that they all performed nearly identical.

\paragraph{Masked particle modelling approach}
For masked particle modelling, we use a slightly modified version of the set-to-set flow matching model (SSFM)~\cite{mpm2}. We have to modify the original SSFM model since it only pre-trains an unpooled set-to-set transformer encoder, while our probing metric requires a single jet representation for linear- and MLP-probing. We therefore modify the SSFM encoder to output a single pooled token, which is used instead of the full encoded set as conditioning for the noisy inputs of the Diffusion Transformer (DiT)~\cite{dit} decoder. Following the conditioning mechanisms of DiT: token concatenation, cross-attention, and adaptive layer norm, we found that concatenating the encoder token to the decoder inputs yielded the best probe performance (see \cref{app:ablation:ssfm} for details).

We note that the original continuous input NTP~\cite{birk2026omnijetalpha2} and SSFM~\cite{mpm2} studies evaluated downstream performance by applying attention-pooling layers over the full sequence of latent tokens rather than extracting a single global jet embedding. As a result, our baseline metrics are not directly comparable. Moreover, since the information carried on a per-token level may be uneven across pre-training objectives, relative fine-tuning performance could differ if all latent tokens were utilised.

\paragraph{Self-distillation approach}
We also compare against a self-distillation approach following the RINO model~\cite{rino} to show that our representation-based pre-training method also improves upon other representation-based pre-training methods and not just reconstruction-based pre-training methods.

\begin{table}[t]
    \centering
    \caption{JetClass classification accuracy for different self-supervised pre-training methods using a frozen backbone with either a linear or MLP probe. For each baseline family, we report the highest probe accuracy.}
    \label{tab:ssl_comparison}
    \begin{tabular}{lcc}
        \toprule
        Method & Linear Probe & MLP Probe \\
        \midrule
        Autoregressive & 0.430 & 0.554 \\
        MPM (SSFM) & 0.635 & 0.715 \\
        Self-distillation (RINO) & 0.533 & 0.648 \\
        \midrule
        LeJEPA (Transformer) & 0.762 & 0.792 \\
        LeJEPA (\lslim) & \textbf{0.774} & \textbf{0.800} \\
        \bottomrule
    \end{tabular}
\end{table}

\paragraph{Comparison}
In \cref{tab:ssl_comparison}, we report the linear- and MLP-probe accuracies on JetClass for the best-performing model configuration within each SSL category, alongside both the standard Transformer and \lslim variants of LeJEPA. Among the baselines, the autoregressive approach yields the lowest linear-probe accuracy. While the self-distillation baseline (RINO) improves probe performance over the autoregressive baseline, the modified SSFM model achieves the best linear and MLP probe among the tested baseline models. 

Even when restricted to a standard non-invariant Transformer backbone, LeJEPA outperforms all baselines. Incorporating explicit Lorentz symmetry via \lslim yields a further improvement on both the linear and MLP probes, demonstrating that Lorentz-invariant architectures and the LeJEPA formulation are complementary in representation learning. For fine-tuning the LeJEPA-based models, we only use the JetCLR augmentation pipeline including Lorentz-transformations, since we have established that this was working best in \cref{sec:res:views}.

\subsection{\label{sec:res:downstream} Downstream performance}
Finally we compared how our LeJEPA-based pre-training approach compares when doing full parameter fine-tuning on limited data sizes. For all fine-tuning results we re-initialize the probe module with random parameters, since in a purely self-supervised setting without pre-training labels the probe cannot be trained. To stabilise fine-tuning we use Linear-Probe fine-tuning (LPFT)~\cite{lpft}, in which the backbone is frozen until the linear classifier is converged after which the backbone parameters are unfrozen and the full model parameters trained. Additionally we use layer-wise learning rate decay (LLRD)~\cite{llrd} in which the learning rate used for each transformer block differs according to a geometric sequence with the smallest learning rate used by the initial block and the largest learning rate used by the final block. For frozen backbone fine-tuning we use full batch L-BFGS similar to our report probes in previous sections. We provide additional information on hyper-parameters in \cref{app:hyperparam}. We plot the mean and standard deviation obtained from the same pre-trained model but fine-tuned across five different seeds.

\begin{figure}[thb]
    \centering
    \includegraphics[width=1\linewidth]{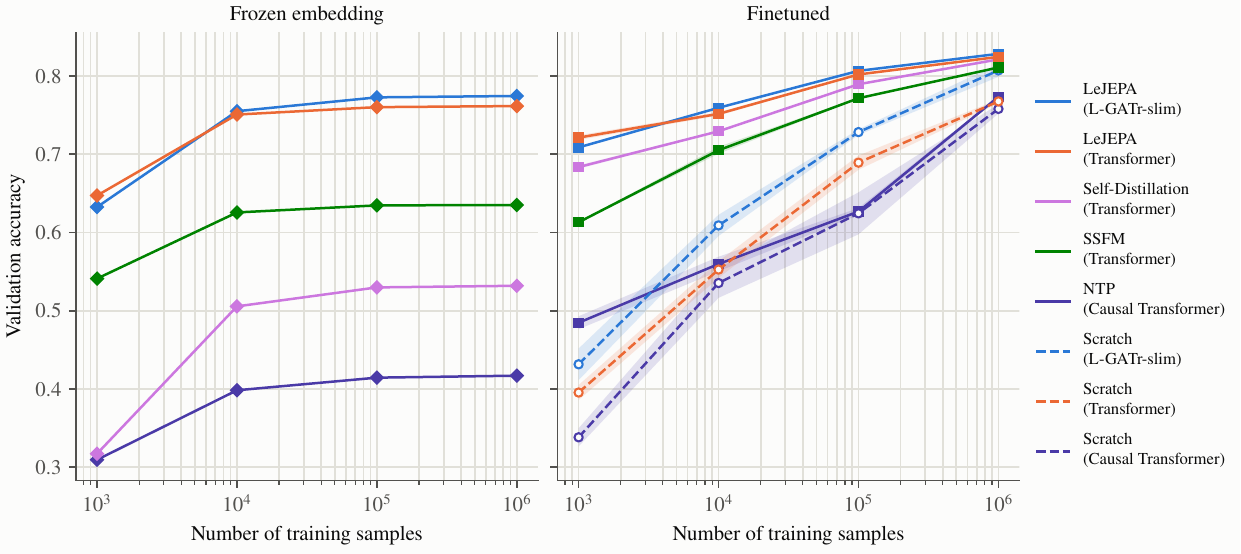}
    \caption{Left: In domain multi-class validation accuracy obtained from linear probes on frozen backbones. Right: In domain multi-class validation accuracy. The solid lines indicate fine-tuning of pre-trained models. Dashed lines are the same architectures trained from scratch. The x-axis denotes the number of used training samples, with the y-axis denotes the measured 10-class validation accuracy, while the x-axis denotes the number of used training samples.}
    \label{fig:jetclass_fine-tune}
\end{figure}

\paragraph{\label{sec:res:jetclass}JetClass fine-tuning}
Here we explore how well our models perform when fine-tuned in domain on JetClass. In \cref{fig:jetclass_fine-tune} we see that our LeJEPA models are outperforming the other pre-training methods especially in the frozen embedding setting with the gap closing the fine-tuned setting. We observe that while SSFM achieves better probe performance than the self-distillation approach, the ordering is switched with self-distillation outperforming SSFM evaluated under full fine-tuning performance. This indicates that while linear-probe performance does correlate with fine-tuned performance the correspondance is not one-to-one. We note that all pre-training methods outperforms their respective backbones trained from scratch, with the LeJEPA models achieving the best performance for all evaluated data sizes. The LeJEPA pre-trained \lslim achieves the best performance for all data samples except the smallest case using 1000 samples, in which the LeJEPA pre-trained transformer outperforms it. This indicates that the benefits, which can be achieved from further developing clever architectures can be combined with pre-training.

\begin{figure}[thb]
    \centering
    \includegraphics[width=1\linewidth]{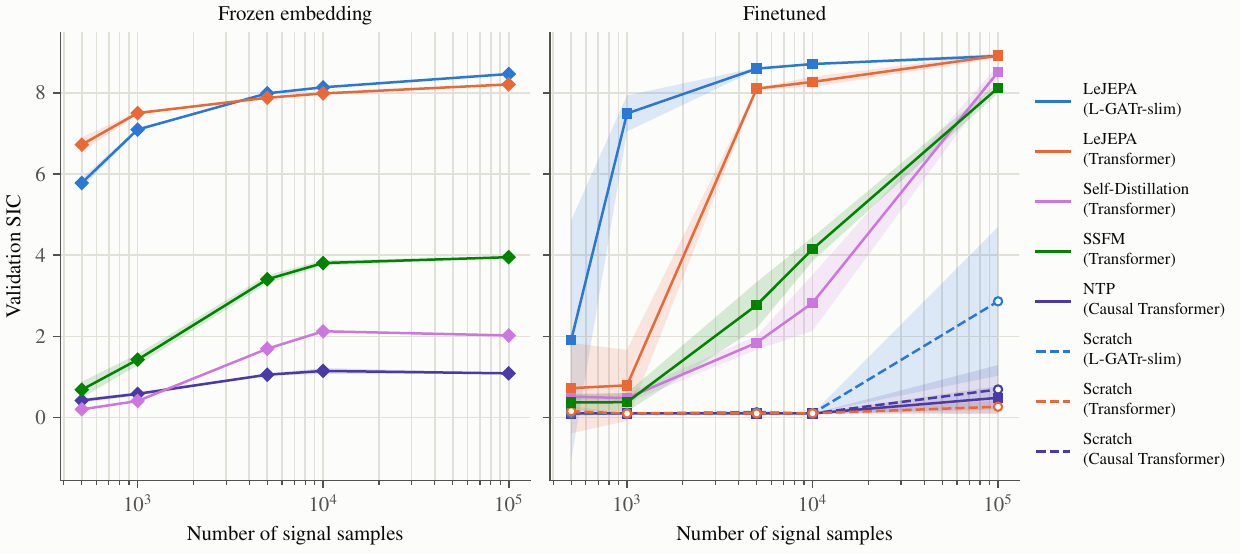}
    \caption{Left: Linear probes on frozen backbones. Right: Solid lines indicate fine-tuning of pre-trained models. Dashed lines are the same architectures trained from scratch. The x-axis denotes the number of signal samples injected one of the background samples, while the y-axis denotes the significance improvement metric (SIC)~\cite{sic}.}
    \label{fig:cwola}
\end{figure}

\paragraph{\label{sec:res:cwola} Weakly supervised anomaly detection}
For the anomaly detection setup, we reuse the JetClass dataset, and change to the Classification Without Labels (CWoLa)~\cite{cwola} setup. According to the CWoLa principle, the optimal classifier between two samples consisting of mixtures of signal and background in which only the proportion of signal and background differs the optimal classifier trained to distinguish between these two samples is equivalent to the optimal classifier between signal and background~\cite{cwola}. To create this setup, we consider two background samples each consisting of \num{500000} QCD jets, where top initiated jets are injected into one sample as signals. In \cref{fig:cwola}, we report the significance improvement (SIC)~\cite{sic} defined as the signal efficiency (true positive rate (TPR), \(\varepsilon_{\text{TPR}}\)) divided by the square root of the background efficiency (false positive rate (FPR), \(\varepsilon_\text{FPR}\)) evaluated at a background efficiency of \(1\unit{\percent}\):
\begin{equation}
    \left. \frac{\varepsilon_{\text{TPR}}}{\sqrt{\varepsilon_\text{FPR}}} \right\vert_{\varepsilon_\text{FPR} = 0.01}.
\end{equation}
Here, we again observe that the LeJEPA-based pre-training approach outperforms the reference methods in both the full fine-tuning example and in the linear probe setting. Additionally, it is seen that \lslim outperforms the transformer model for all signal strengths in the fully fine-tuned setting. In the lowest-signal regime linear probing on the frozen LeJEPA representations outperforms full fine-tuning, which likely suffers from over-fitting as observed through the high high variance across different seeds.

\begin{figure}[thb]
    \centering
    \includegraphics[width=1\linewidth]{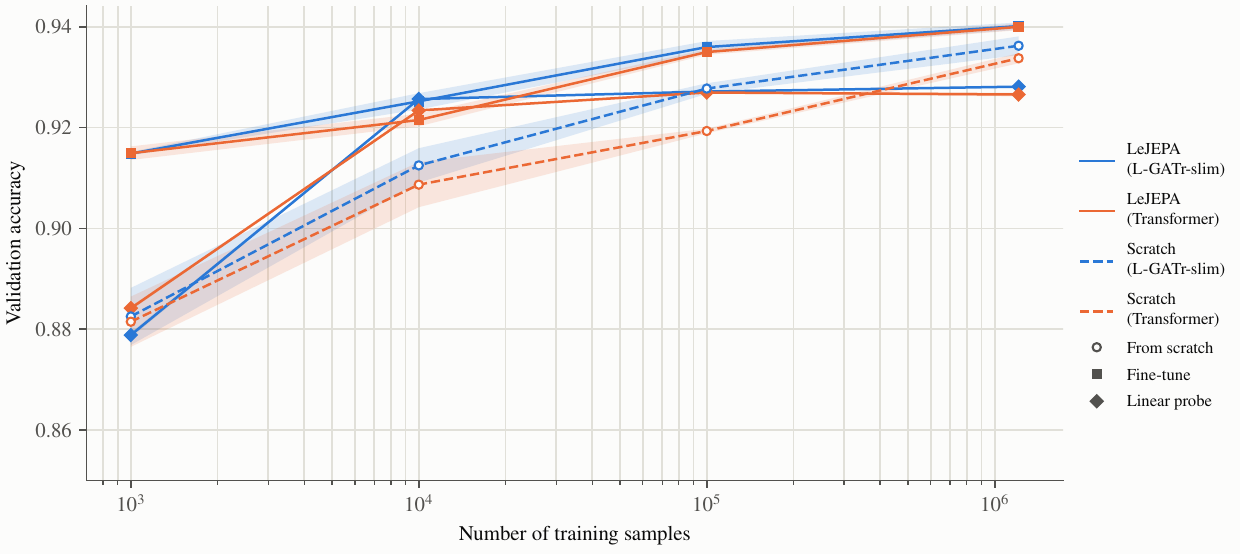}
    \caption{Transfer learning on top tagging dataset. Solid lines represent pre-trained models, while dashed lines are from scratch. Additionally, squares denote full finetuning, while diamonds denote frozen backbone performance. The circles and dotted lines indicates from scratch training. The x-axis denotes the number of training samples, while the y-axis denotes the binary classification accuracy.}
    \label{fig:toptag}
\end{figure}

\paragraph{\label{sec:res:toptag} Top tagging transfer learning}
Finally, we explore how our models transfer to a dataset using different detector settings. Since the top tagging dataset only includes kinematic variables, we redo the pre-training on JetClass using only the kinematic features. As such, we only include the performance of the LeJEPA-based models in this evaluation due to computational considerations. In \cref{fig:toptag} we find that our LeJEPA-based pre-training transfers to a different dataset, since the fine-tuned models outperform the from-scratch baselines. Additionally, we observe the pre-trained \lslim slightly outperforms the pre-trained transformer. This showcases that advances in architectures can be combined with self-supervised pre-training.

\section{Conclusion}
\label{sec:conclusion}
In this work, we bridged representation-based self-supervised learning and exact spacetime symmetries for jet physics by combining the LeJEPA framework with the Lorentz-invariant \lslim architecture. Through systematic comparisons against a standard Transformer baseline, we showed that enforcing Lorentz invariance in the learned latent representation is a decisive driver of representation quality: without \(\text{SO}^+(1,3)\) invariance enforced either structurally or through global kinematic augmentations, linear- and MLP-probe accuracies degrade severely. 

The LeJEPA framework outperforms reconstruction- and self-distillation-based SSL paradigms in HEP both when comparing probe performance and fully fine-tuned performance. Because our pre-training objective requires no ground-truth labels, this framework provides a promising blueprint for training physics-informed foundation models directly on unlabelled experimental collider data.

Due to computational constraints we were restricted to only explore the LeJEPA paradigm for relatively small models. This represents a limitation of the results represented in this work. As such it is still an open question how well the framework scales, both in training data and model expressivity. 

\section*{Acknowledgements}
We would like to thank Antoine Petitjean and Sebastian Pitz for stimulating discussions providing insights into Lorentz-invariant models. We would also like to thank L\'{e}o Nicollier for many valuable discussions regarding LeJEPA-derived models, in particular SphereJEPA. Finally, we would like to thank Pradyun Hebbar and Chris Scheulen for valuable comments on the manuscript.

This work is supported by the European Union's Horizon Europe research and innovation program under the Marie Sk{\l}odowska-Curie grant agreement No 101168829 called ``Challenging AI with Challenges from Physics: How to solve fundamental problems in Physics by AI and vice versa (AIPHY)''.

The training and evaluation of the models developed in this work were performed at the University of Geneva using the Baobab high-performance computing cluster.

\clearpage

\bibliographystyle{iopart-num}
\bibliography{main}

\clearpage

\appendix
\crefalias{section}{appendix}

\section{\label{app:hyperparam} Hyperparameters}
Both the standard transformer and \lslim backbones are configured in similar manners to Reference~\cite{lgatrslim} which matches the total number of scalar channels and vector channels of \lslim with the total number of channels in the transformer. After the primary constituent encoder, per-constituent representations are projected to \(\RR^d\) and aggregated via two standard Transformer layers with a prepended learnable \textsc{cls} token. For LeJEPA pre-training the Epps-Pulley integral in \cref{eq:sigreg_expectation} is evaluated numerically via the composite trapezoidal rule using 17 equally spaced knots on \(\lrb{-3, 3}\). For pre-training a summary of the architectural and optimization hyperparameters is provided in \cref{tab:pt_hyperparams}.

\begin{table}[h]
    \centering
    \caption{Default architectural and pre-training hyperparameters for the standard Transformer and \lslim backbones.}
    \label{tab:pt_hyperparams}
    \begin{tabular}{lc}
        \toprule
        Hyperparameter & Transformer [\lslim] \\
        \midrule
        Embedding dim \(\lrp{d_s, d_v}\) & \((\num{128}, \num{0})\) [\((\num{96}, \num{32})\)] \\
        Encoder layers & \(6\) \\
        Pooling layers & \(\num{2}\) \\
        Register [Spurion] tokens & \(\num{3}\) \\
        LeJEPA weight \(\lambda\) & \(\num{0.02}\) \\
        SIGReg dimension & \(\num{16}\) \\
        Random projections & \(\num{256}\)  \\
        Epps-Pulley knots &  \(\num{17}\) \\
        Integration interval & \(\lrb{\num{-3}, \num{3}}\) \\
        Batch size \(\abs{B}\) & \(\num{1000}\) \\
        JetCLR views \(\abs{V}\) & \(\num{4}\) \\
        RINO views \(\abs{V}\) & \(\num{8}\) \\
        Optimizer & AdamW~\cite{adamw2018} \\
        Learning rate & \(\num{0.0003}\) \\
        Weight decay & \(\num{0.01}\) \\
        Scheduler & Cosine Annealing~\cite{cosineanneal} \\
        Total steps & \num{200000} \\
        \bottomrule
    \end{tabular}
\end{table}

In \cref{tab:ft_hyperparams} we list the hyperparameters for our finetuning and scratch procedure. The hyper-parameters for LPFT~\cite{lpft} and LLRD~\cite{llrd} are only applied to the finetuning models, the from scratch models are trained using normal optimisation of all the parameters. We generally use the same batch size as pre-training (1000) except for the case where we train on 10000 or fewer total samples, in which case we reduce the batch size to 100. For the frozen embeddings we use full batch L-BFGS~\cite{nocedal1980lbfgs}, while for the finetuning we use the Lion optimizer~\cite{lion2023}, since it was chosen as the optimizer for jet tagging of \lslim~\cite{lgatrslim}.
\begin{table}[h]
    \centering
    \caption{Fine-tuning hyperparameters for the standard Transformer and \lslim backbones.}
    \label{tab:ft_hyperparams}
    \begin{tabular}{lc}
        \toprule
        Hyperparameter & Value(s) \\
        \midrule
        Batch size \(\abs{B}\) & \(\num{100} / \num{1000}\) \\
        Optimizer & L-BFGS~\cite{nocedal1980lbfgs} / Lion~\cite{lion2023} \\
        Learning rate & \(\num{0.0003}\) \\
        Weight decay & \(\num{0.00}\) \\
        Scheduler & Cosine Annealing~\cite{cosineanneal} \\
        Total steps & \num{40000} \\
        Frozen steps & \num{12000} \\
        Layer-wise LR Decay & \num{0.65} \\
        \bottomrule
    \end{tabular}
\end{table}

\section{\label{app:ablation} Ablation studies}

\subsection{\label{app:ablation:views} Views ablation}
In this section we study the impact of two design choices in the LeJEPA view-creation pipeline. Firstly we ablate the number of total views and secondly we ablate the complexity of the applied augmentations.

\begin{figure}[t]
    \centering
    \includegraphics[width=1\linewidth]{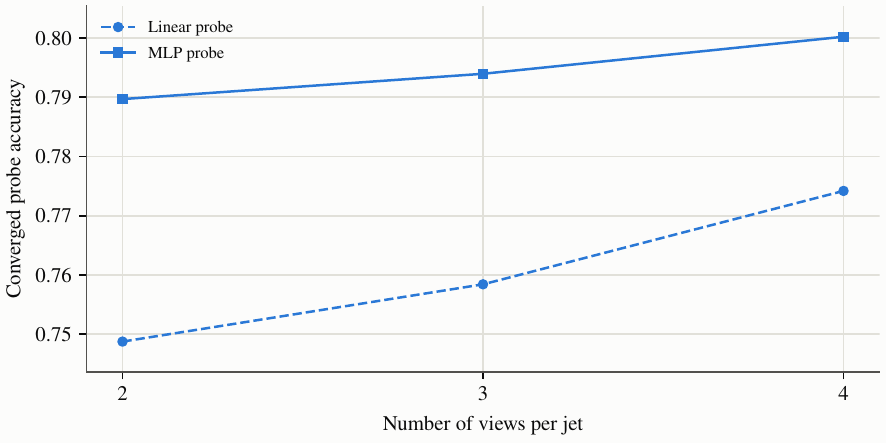}
    \caption{Graph showing the linear (dashed)- and MLP (solid)-probe performance as a function of the number of views used in the LeJEPA training objective, for the \lslim model. We see that the probe performance is monotonically increasing with the number of views.}
    \label{fig:number_views}
\end{figure}

\paragraph{Number of views}
Here we show that using as little as two views, which consists of the original jet and a single augmentation. Our method achieves very strong probe performance and this performance increases with additional views up to our deployed configuration using 4 total views as shown in \cref{fig:number_views}. Due to hardware limitations we did not explore further increasing the views.

\paragraph{Augmentation complexity}
The JetCLR and RINO based augmentation pipelines involve various physically motivated augmentations, we here seek a minimal set of simple augmentations which reach comparable results. In \cref{sec:res:views} we including \(\text{SO}^+(1,3)\) yields substantial performance improvement when done in conjunction with the JetCLR or RINO augmentations, however in \cref{tab:ablation_minimal} we show that \(\text{SO}^+(1,3)\) alone is not sufficient. Additionally we find that purely applying a masking augmentation is likewise insufficient. However, applying both in combination surpasses all the reference SSL methods, showcasing that the LeJEPA framework is competitive whilst using only a minimal augmentation strategy consisting purely of random masking and Lorentz transformations as shown in \cref{tab:ablation_minimal}, but that including a full set of physics motivated augmentations further improves performance.

\begin{table}[t]
    \centering
    \caption{Ablation of minimal view-generation strategies on JetClass. The minimal setup, which consists of purely particle masking and global \(\text{SO}^+(1,3)\) Lorentz transformations, are compared against the full JetCLR augmentation pipeline described in \cref{sec:augmentations}}.
    \label{tab:ablation_minimal}
    \begin{tabular}{llcccc}
        \toprule
        Probe & Model & \(\text{SO}^+(1,3)\) & Masking & Masking                  & Full JetCLR \\
              &       &                      &         & \(+\, \text{SO}^+(1,3)\) & \(+\, \text{SO}^+(1,3)\) \\
        \midrule
        Linear & \lslim      & 0.412 & 0.469 & 0.709 & 0.774 \\
               & Transformer & 0.360 & 0.295 & 0.622 & 0.762 \\
        \midrule
        MLP    & \lslim      & 0.485 & 0.583 & 0.766 & 0.800 \\
               & Transformer & 0.409 & 0.379 & 0.687 & 0.792 \\
        \bottomrule
    \end{tabular}
\end{table}

\subsection{\label{app:ablation:sigreg} Target distributions in SIGReg}
As described in \cref{sec:lejepa}, SIGReg prevents representation collapse by matching the empirical characteristic function \(\hat{\phi}_{\lra{Z, a}}(s)\) of the projected embeddings along random unit directions \(a \in S^{d-1}\) to the one-dimensional marginal characteristic function \(\phi(s) = \EE\lrb{\exp\lrp{\imag s \lra{Y, a}}}\) of an isotropic target \(Y \in \RR^d\). We compare three isotropic target distributions:
\begin{enumerate}
    \item \textbf{Standard multivariate normal}~\cite{lejepa},
    
    \(Y \sim \mathcal{N}(0, \II_d)\), whose one-dimensional marginal along any \(a \in S^{d-1}\) is \(\mathcal{N}(0, 1)\) with characteristic function
    \begin{equation}
        \phi_\text{Gaussian}(s) = \exp \lrp{-\frac{s^2}{2}}.
    \end{equation}
    \item \textbf{Uniform hyperspherical distribution}~\cite{spherejepa},
    
    \(Y \sim \mathcal{U}( S^{d-1})\), with characteristic function given by the normalized Bessel function of the first kind,
    \begin{equation}
        \phi_\text{Sphere}(s) = \Gamma\lrp{\frac{d}{2}} \lrp{\frac{2}{\abs{s}}}^{\frac{d}{2}-1} J_{\frac{d}{2}-1}\lrp{\abs{s}}.
    \end{equation}
    \item \textbf{Symmetric multivariate Laplace distribution}~\cite{kerjepa}, defined by the heavier-tailed marginal characteristic function
    \begin{equation}
        \phi_\text{Laplace}(s) = \lrp{\frac{1}{1 + s^2}}^{\frac{d+1}{2}},
    \end{equation}
    where \(d\) denotes the dimension of the embedded space.
    This characteristic function matches a probability density proportional to
    \begin{equation}
        p(x) \propto \exp \lrp{-\norm{x}_2}
    \end{equation}
\end{enumerate}
As shown in \cref{tab:ablation_sigreg}, we find that regularizing toward the multivariate Laplace distribution yields the highest probe accuracy for both architectures.

\begin{table}[h]
    \centering
    \caption{Comparison of target distributions for the SIGReg regularizer in LeJEPA.}
    \label{tab:ablation_sigreg}
    \begin{tabular}{lcc}
        \toprule
        Target distribution & \lslim & Transformer \\
        \midrule
        Linear: \\
        -- Multivariate Normal & 0.723 & 0.736 \\
        -- Uniform Hyperspherical & 0.582 & 0.467 \\
        -- Multivariate Laplace & \textbf{0.774} & \textbf{0.762} \\
        \midrule
        MLP: \\
        -- Multivariate Normal & 0.780 & 0.779 \\
        -- Uniform Hyperspherical & 0.671 & 0.562 \\
        -- Multivariate Laplace & \textbf{0.800} & \textbf{0.792} \\
        \bottomrule
    \end{tabular}
\end{table}

\subsection{\label{app:ablation:ar} Autoregressive feature prediction subsets}
For the autoregressive baseline we match the transformer backbone  (\cref{sec:res:ssl}) to the other transformers as much as possible, however since the model needs to be causal for the training objective to work, we use causal attention. Additionally since an autoregressive model predicts on a sequence we provide Rotary Position Encodings to break the permutation equivariance of a transformer. The loss is the average of each constituent token-loss, which as described in the main text combines discretised kinematics (trained via cross-entropy over VQ-VAE~\cite{vqvae} codebook indices), continuous impact parameters \((z_0, \sigma(z_0), d_0, \sigma(d_0))\) (trained via \(L_2\) regression), and 8-dimensional identification ID one-hot vectors (trained via cross-entropy). In \cref{tab:ablation_ar}, we evaluate the effect of restricting the pre-training loss to subsets of these variables. We find here that the differences between all the different tried objectives are very small, but ultimately predicting only the VQ-VAE tokens yield the best linear and MLP probe performances.

\begin{table}[h]
    \centering
    \caption{Ablation over predicted feature subsets for the autoregressive baseline on JetClass.}
    \label{tab:ablation_ar}
    \begin{tabular}{lcc}
        \toprule
        Predicted Features & Linear Probe & MLP Probe \\
        \midrule
        Kinematics only (VQ-VAE) & \textbf{0.430} & \textbf{0.554} \\
        Kinematics + ID & 0.414 & 0.548 \\
        Kinematics + Impact parameters & 0.416 & 0.544 \\
        All & 0.422 & 0.550 \\
        \bottomrule
    \end{tabular}
\end{table}

\subsection{\label{app:ablation:ssfm} Conditioning mechanisms in SSFM}
To adapt the set-to-set flow matching (SSFM) model~\cite{mpm2} for representation learning (\cref{sec:res:ssl}), we pool the encoder outputs into a single latent jet token \(z \in \RR^d\) and condition the Diffusion Transformer (DiT)~\cite{dit} decoder on \(z\) alongside the flow-matching time-step \(t \in \lrb{0, 1}\). Following Reference~\cite{dit}, we evaluate three conditioning strategies: (i) cross-attention, where decoder blocks attend to \(z\) via an auxiliary multi-head cross-attention layer; (ii) adaptive layer norm, where \(z\) and the time-step embedding modulate the scale and shift parameters of the decoder layer normalization layers; and (iii) token concatenation (memory token), where \(z\) is concatenated to the noisy constituent set used as input for the decoder. 

\begin{table}[h]
    \centering
    \caption{Comparison of DiT decoder conditioning mechanisms for the modified SSFM baseline on JetClass.}
    \label{tab:ablation_ssfm}
    \begin{tabular}{lcc}
        \toprule
        SSFM Conditioning & Linear Probe & MLP Probe \\
        \midrule
        Cross-attention & 0.467 & 0.550 \\
        Adaptive layer norm & 0.613 & 0.697 \\
        Concatenated token & \textbf{0.635} & \textbf{0.715} \\
        \bottomrule
    \end{tabular}
\end{table}

As shown in \cref{tab:ablation_ssfm}, we find that using cross-attention as the conditioning mechanism performed the worst, while using a concatenated memory-token or AdaLN drastically outperforms the cross-attention mechanism, with the concatenated memory-token slightly outperforming AdaLN. We therefore use token concatenation SSFM version as our MPM baseline in \cref{tab:ssl_comparison}, we note that this is conditioning mechanism differs from the findings of DiT~\cite{dit}, where AdaLN performed best for generative conditioning and the concatenated token being the worst.

\clearpage